\documentclass[fleqn,usenatbib]{mnras}

\usepackage{newtxtext,newtxmath}

\usepackage[T1]{fontenc}

\DeclareRobustCommand{\VAN}[3]{#2}
\let\VANthebibliography\thebibliography
\def\thebibliography{\DeclareRobustCommand{\VAN}[3]{##3}\VANthebibliography}

\usepackage{graphicx}	
\usepackage{amsmath}	
\usepackage{siunitx}
\usepackage{pifont}
\newcommand{\angstrom}{\text{\normalfont\AA}}

\title[Population III Stars in the Early Universe]{The Challenges of Identifying Population III Stars in the Early Universe}

\author[H. Katz] {Harley Katz$^{1}$\thanks{E-mail:
  \href{mailto:harley.katz@physics.ox.ac.uk}{harley.katz@physics.ox.ac.uk}}, Taysun Kimm$^2$, Richard S. Ellis$^3$, Julien Devriendt$^1$, \& Adrianne Slyz$^1$ 
  \\
  $^1$Sub-department of Astrophysics, University of Oxford,
   Keble Road, Oxford OX1 3RH, UK \\
  $^2$Department of Astronomy, Yonsei University, 50 Yonsei-ro,
  Seodaemun-gu, Seoul 03722, Republic of Korea \\  
  $^3$Department of Physics and Astronomy, University College London, Gower Street, London WC1E 6BT, UK \\
  }

\date{Accepted XXX. Received YYY; in original form ZZZ}

\pubyear{2021}

\begin{document}
\label{firstpage}
\pagerange{\pageref{firstpage}--\pageref{lastpage}}
\maketitle

\begin{abstract}
The recent launch of JWST has enabled the exciting prospect of detecting the first generation of metal-free, Population III (Pop.~III) stars. Determining the emission line signatures that robustly signify the presence of Pop.~III stars against other possible contaminants represents a key challenge for interpreting JWST data. To this end, we run high-resolution (sub-pc) cosmological radiation hydrodynamics simulations of the region around a dwarf galaxy at $z\geq10$ to predict the emission line signatures of the Pop.~III/Pop.~II transition. We show that the absence of metal emission lines is a poor diagnostic of Pop.~III stars because metal-enriched galaxies in our simulation can maintain low [O{\small III}]~5007\angstrom~emission that may be undetectable due to sensitivity limits. Combining spectral hardness probes (e.g. He{\small II}~1640\angstrom/H$\alpha$) with metallicity diagnostics is more likely to probe the presence of metal-free stars, although contamination from Wolf-Rayet stars, X-ray binaries, or black holes may be important. The hard emission from Pop.~III galaxies fades fast due to the short stellar lifetimes of massive Pop.~III stars, which could further inhibit detection. Similarly, Pop.~III stars may be detectable after they evolve off the main-sequence due to the cooling radiation from nebular gas or a supernova remnant; however, these signatures are also short-lived (i.e. few Myr), and contaminants such as flickering black holes might confuse this diagnostic. While JWST will provide a unique opportunity to spectroscopically probe the nature of the earliest galaxies, both the short timescales associated with pristine systems and ambiguities in interpreting key diagnostic emission lines may hinder progress. Special care will be needed before claiming the discovery of systems with pure Pop.~III stars. 
\end{abstract}

\begin{keywords}
 galaxies: evolution, galaxies: formation, galaxies: high-redshift, stars: formation, stars: Population III
\end{keywords}



\section{Introduction}
One of the most compelling scientific goals of the James Webb Space Telescope (JWST) is either detecting or placing strong constraints on the properties of the first generation of metal-free, Population~III (Pop.~III) stars \citep{Gardner2006}. Very little is known about their properties, such as when they started forming, when they stopped forming, their initial mass function (IMF), or metal yields because to date, there has yet to be a robust detection of Pop.~III stars \textemdash~ despite there being speculation about a few peculiar objects \citep[e.g.][]{Sobral2015,Vanzella2020,Welch2022}. The constraints we have on their characteristics are either derived from high-resolution numerical simulations \citep[e.g.][]{Abel2002,Bromm2002,Stacy2010,Greif2011,Hirano2014,Hosokawa2016} or from stellar archaeology around the Milky Way \citep[e.g.][]{Beers2005,Frebel2007,Karlsson2013}. Nevertheless, this is hopefully set to change with the recent launch of JWST or future facilities such as HARMONI on the ELT \citep{Grisdale2021}.

Even with optimistic assumptions on the Pop.~III IMF and the redshift to which some gas in the Universe can remain pristine, it is unlikely that individual metal-free stars will be detectable with JWST \citep[e.g.][]{Zackrisson2011,Rydberg2013,Schauer2020}. Although, prospects are better if Pop.~III stars form in large groups and a significant amount of observing time is spent on lensing clusters \citep[e.g.][]{Stiavelli2010,Pawlik2011,Jeon2019,Vikaeus2022}. Direct constraints on the Pop.~III IMF may also arise from observing their SN, if they happen to be particularly bright, or result in a gamma-ray burst \citep{Lazar2022}.

One of the primary issues with robustly detecting Pop.~III stars is determining the spectral signatures that differentiate them from other classes of objects that form in the early Universe. Such contaminants could include direct collapse black holes or a second generation of already metal enriched stars \citep{Nakajima2022}. Due to their possible higher masses and metal-free nature, the spectra of Pop.~III stars are expected to be considerably harder than their Pop.~II counterparts \citep[e.g.][]{Schaerer2002}. For this reason, it has been postulated that spectral signatures from high-energy ionization states \citep[e.g. He{\small II},][]{Oh2001,Tumlinson2001} represent one possible indirect signature of Pop.~III stars. However, strong He{\small II} lines by themselves are not a definitive signature of metal-free stars because other sources, such as Wolf-Rayet stars, X-ray binaries, or black holes can similarly produce strong He{\small II} emission \citep[e.g.][]{Erb2010,Inoue2011,Schaerer2019,Nakajima2022}. Pop.~III stars are also predicted to excite strong hydrogen emission lines such a Ly$\alpha$, H$\alpha$, and H$\beta$, although the former is unlikely to be detectable due to the optically thick IGM at high redshifts \citep{Gunn1965,Inoue2014}. Another possible contaminant are galaxies that have metals but the metal emission lines are simply too weak to be detected. 

Given the recent launch of JWST, it is timely to revisit the spectral signatures of Pop.~III stars in the context of fully coupled radiation hydrodynamics simulations that resolve much of the physics driving emission lines. This provides a complementary view to the various photoionization models that have been analysed on this topic \citep[e.g.][]{Inoue2011,Nakajima2022}.

\begin{figure}
\centerline{\includegraphics[width=3.31in,keepaspectratio,trim={0 0.0cm 0cm 0cm},clip]{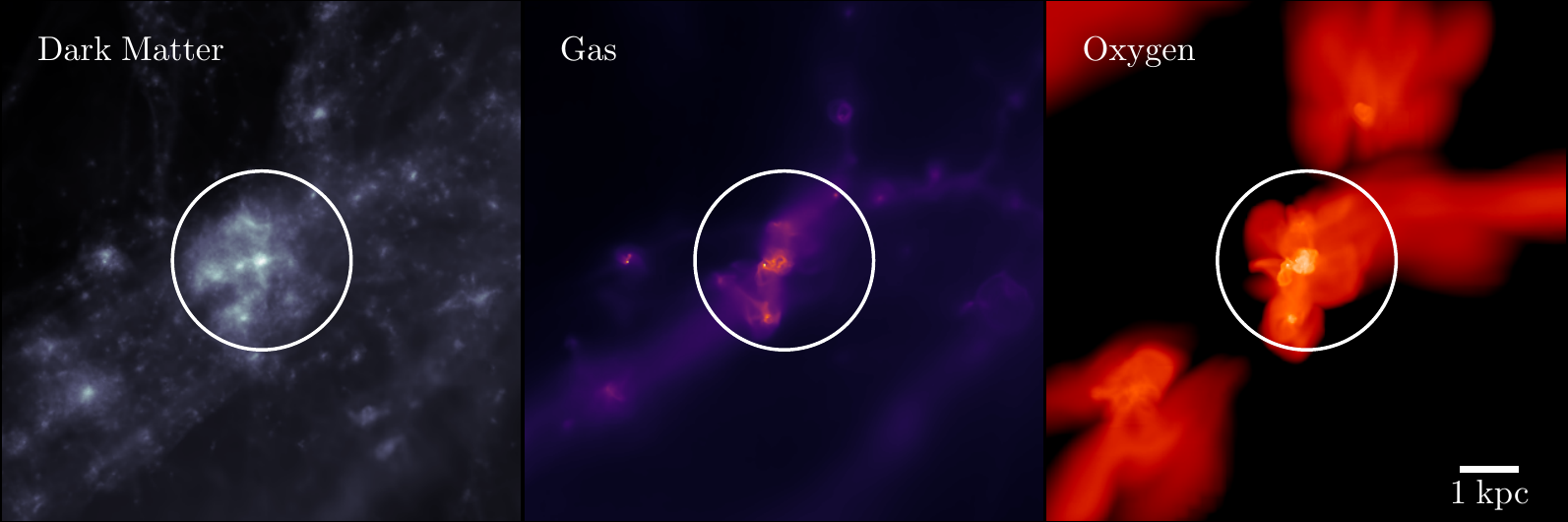}}
\centerline{\includegraphics[width=3.31in,keepaspectratio,trim={0 0.0cm 0cm 0cm},clip]{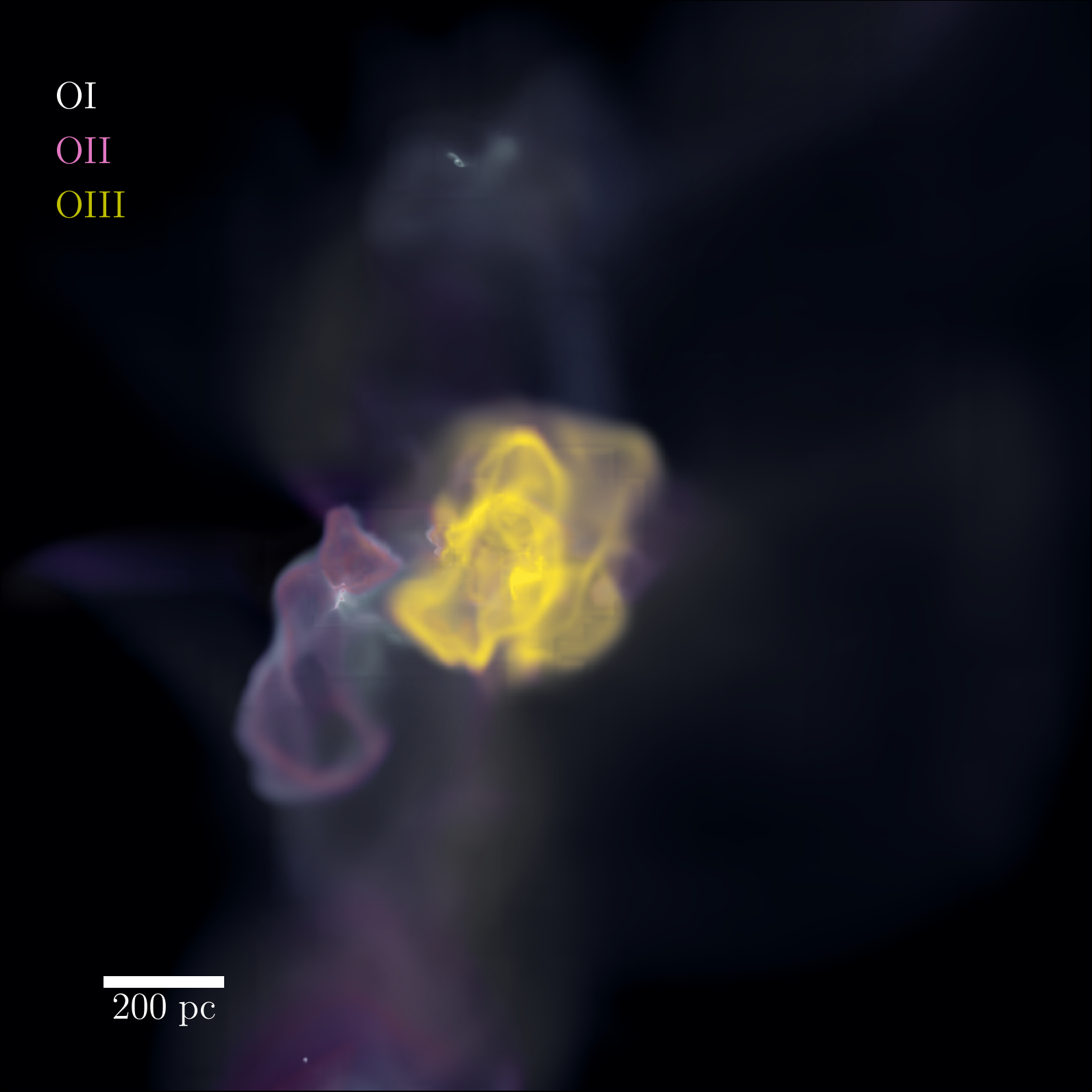}}
\caption{(Top) Maps of dark matter, gas, and oxygen in the simulation at $z=10$. The white circle shows the virial radius (1.72~kpc) of the most massive halo in the simulation. (Bottom) Map of O{\small I}, O{\small II}, and O{\small III} in the central regions of the most massive halo.}
\label{hero}
\end{figure}

\section{Methods}
For this work, we employ high-resolution cosmological radiation hydrodynamics simulations run with the {\small RAMSES-RTZ} code \citep{Katz2022}, an extension of the {\small RAMSES-RT} code \citep{Teyssier2002,Rosdahl2013,Rosdahl2015} that includes modules for radiation-coupled H$_2$ \citep{Katz2017} and metal chemistry \citep{Katz2022}. The underlying physics closely follows that presented in \cite{Kimm2017} and here we briefly highlight the features included in the simulation with a particular emphasis on the changes made for these runs.

The simulations follow gravity, hydrodynamics, radiation transfer, and various chemistry, cooling, and heating processes. The radiation is followed in 8 frequency bins (see Table~2 in \citealt{Kimm2017}) using the M1 method \citep{Levermore1984}. To reduce the computational cost of the RT, we employ a reduced speed of light approximation, adopting $c_{\rm sim}=0.01c$. The radiation is coupled to the gas via photoionization, photoheating, and radiation pressure (both direct UV and multi-scattered IR). Similar to \cite{Kimm2017} we follow H{\small I}, H{\small II}, e, He{\small I}, He{\small II}, and He{\small III}, but new for this work, we also follow various ionization states of O, C, N, Fe, Si, S, Ne, and Mg. Atomic data for each metal closely follows that adopted by {\small CLOUDY} v.17 \citep{Ferland2017}. Cooling for primordial species follows the methods presented in \cite{Rosdahl2013,Katz2017} and cooling for metals is calculated at low temperatures ($T<10^4$~K) by computing the equilibrium level populations of certain ions and for high-temperatures ($T\geq10^4$~K) by using look-up tables (see \citealt{Oppenheimer2013,Katz2022}).

Cosmological initial conditions are generated for a halo with mass $\sim3\times10^8$~M$_{\odot}$ at $z=10$ using {\small MUSIC} \citep{Hahn2011}, assuming the following cosmology: $\Omega_{\rm m}=0.311$, $\Omega_{\rm b}=0.045$, $\Omega_{\rm \Lambda}=0.689$, and $h=0.6766$ \citep{planck}. Due to the large computational expense of these simulations, we apply the zoom-in technique and place the bulk of our resolution elements around a single galaxy, ensuring that there is no contamination from low-resolution elements within the virial radius at any point. The dark matter particle mass within the high resolution region is 492~M$_{\odot}$. Throughout the course of the simulation, we allow the adaptive mesh to refine when either the Jeans length is not resolved by at least 8 cells, or the dark matter or gas mass of the cell grows to 8 times its initial value. The minimum cell size is 0.8~pc at $z=10$ and remains constant in co-moving coordinates (resulting in even smaller cell sizes at higher redshifts). 

When the gas becomes dense enough, star particles are allowed to form. This occurs when the turbulent Jeans length is unresolved \citep[see e.g.][]{Kimm2017} and the gas is converging on a local density maximum. If these criteria are satisfied, star formation proceeds in two different modes depending on metallicity. Pop.~III stars can form when the gas metallicity is $<10^{-6}Z_{\odot}$, while Pop.~II stars form at higher metallicities. In the case of Pop.~II stars, particle masses are integer multiples of 500~M$_{\odot}$. The number of star particles formed is drawn from a Poisson distribution with the conversion rate calculated via a Schmidt law \citep{Schmidt1959}. The efficiency of conversion depends on the thermo-turbulent properties of the gas calibrated on high-resolution molecular cloud simulations \citep{Padoan2011,Federrath2012}. Pop.~II star particles are assumed to host stellar populations that follow a Kroupa IMF \citep{Kroupa2001}. In the case of Pop.~III stars, when the star formation criteria are met, we draw individual star particle masses from a stellar IMF following \cite{Wise2012}.

\begin{figure}
\centerline{\includegraphics[width=3.31in,keepaspectratio,trim={0 0.2cm 0cm 0.4cm},clip]{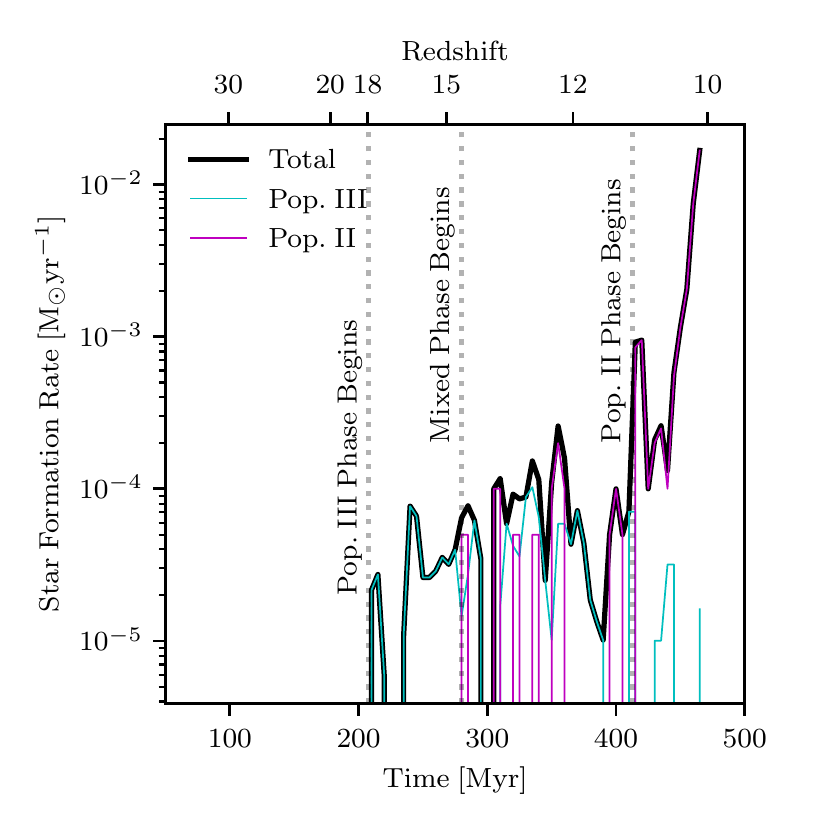}}
\centerline{\includegraphics[width=3.31in,keepaspectratio,trim={0 0.3cm 0cm 0.4cm},clip]{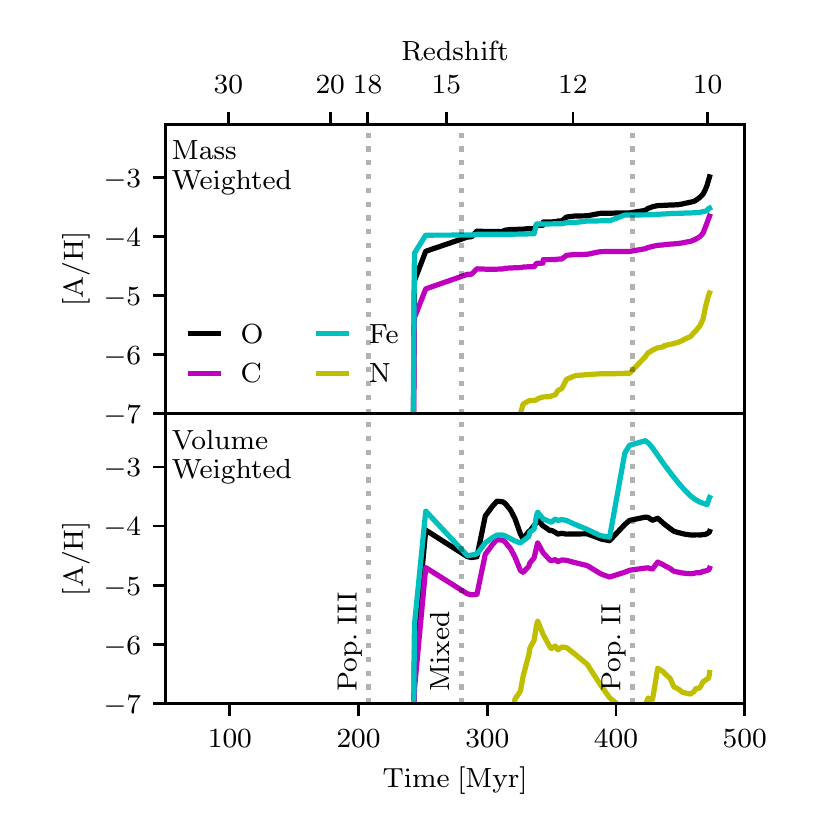}}
\caption{(Top) Star formation rate in the Lagrange volume as a function of time (black) split between the contribution of Pop.~III stars (cyan) and Pop.~II stars (magenta). (Bottom) Mass-weighted or volume-weighted metallicities of individual elements (O, C, Fe, and N) as a function of time.}
\label{sfr_metal_enrich}
\end{figure}

Star particles impact their environment in numerous ways. During the course of their lifetime, they inject photons into their host cells. For Pop.~III stars, we adopt the age and mass dependent SEDs from \cite{Schaerer2002} and for Pop.~II stars, we use the age, mass, and metallicity dependent SEDs from {\small BPASS} v.2.2.1 \citep{Stanway2018}. When Pop.~III stars reach the end of their lifetime, they can either explode via SN or directly collapse to a black hole. The energy per Pop.~III SN and the mass ranges for SN versus direct collapse are adopted from \cite{Wise2012}, which are based on \cite{Woosley1995,Heger2002,Nomoto2006} (see Equations. 22 and 23 of \citealt{Kimm2017}). For Pop.~II stars, we follow both type~II core-collapse SN as well as SNIa. We randomly sample the stellar IMF to determine when to inject core-collapse SN following the age-mass distribution of \cite{Raiteri1996}. When SN occur, momentum is injected into the simulations following the mechanical feedback model of \citep{Kimm2015}. Stellar winds from AGB stars are also included using the method presented in \cite{Agertz2013}.

During all of these energetic feedback processes, metals are released into the gas. In all cases, yields are dependent on stellar mass. For Pop.~III stars, we consider individual SN yields from typical type~II SN \citep{Nomoto2006}, hypernova \citep{Nomoto2006}, and pair-instability SN \citep{Heger2002}. The mass thresholds for each are described in \cite{Kimm2017}. Metal yields for Pop.~II core-collapse SN are adopted from \cite{Portinari1998}, while SNIa yields are from \cite{Seitenzahl2013}. Finally, AGB wind yields are from \cite{Pignatari2016}. We follow the enrichment of O, N, C, Mg, Si, S, Fe, Ne, and Ca.

To analyse the simulations, we use the AMIGA halo finder (AHF) to identify the dark matter haloes \citep{Gill2004,Knollmann2009} using the virial over-density criteria of \cite{Bryan1998}. For each halo, we compute the luminosities of various emission lines. Line emission from primordial species is computed following the fitting functions presented in \cite{Katz2022b}, while metal emission line luminosities are calculated with {\small PyNeb} \citep{Luridiana2015}. Due to the very low metallicities, we neglect the impact of dust on the presented emission line luminosities.

\begin{figure}
\centerline{\includegraphics[width=3.31in,keepaspectratio,trim={0 0.3cm 0cm 0.4cm},clip]{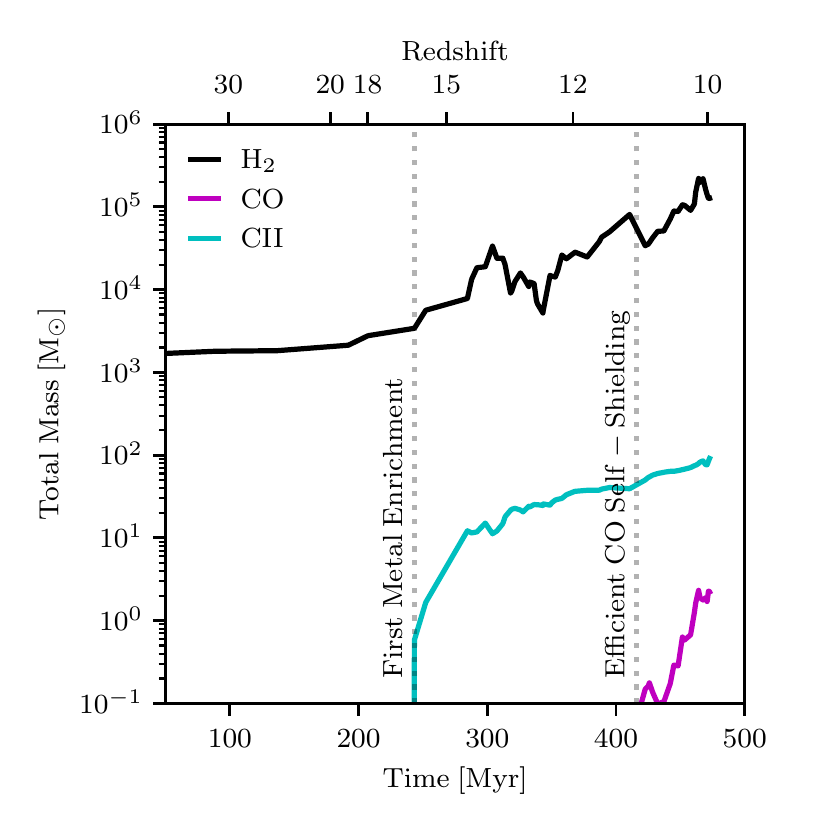}}
\caption{Evolution of the total mass of H$_2$ (black), C{\small II} (cyan), and CO (magenta) within the Lagrange region of the main halo as a function of time.}
\label{molecule}
\end{figure}

\begin{figure*}
\centerline{\includegraphics[scale=1,trim={0 0.0cm 0cm 0cm},clip]{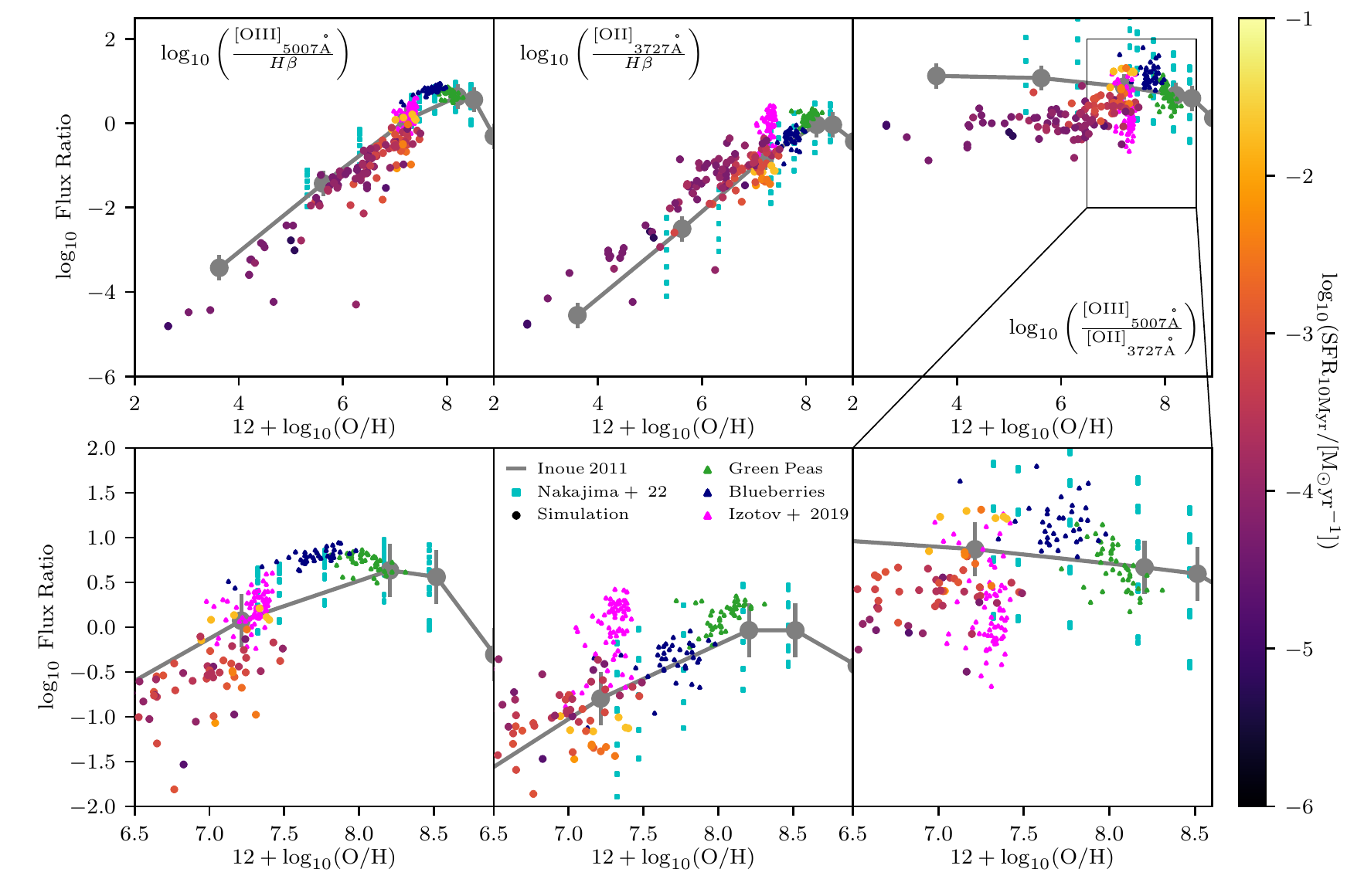}}
\caption{Flux ratios of [O{\small III}]/H$\beta$ (left), [O{\small II}]/H$\beta$ (centre) or [O{\small III}]/[O{\small II}] (right) for the simulated galaxy population in the redshift range $10\leq z\leq16$. The bottom row shows a zoomed in version of the top row. The data points from the simulation are coloured by their 10~Myr-averaged SFR. For comparison, we show results from the {\small CLOUDY} models of \protect\cite{Inoue2011} (grey) and \protect\cite{Nakajima2022} (cyan squares) for primordial stellar populations. Similarly, we show flux ratios for local Green Pea galaxies \protect\citep{Yang2017}, Blueberry galaxies \protect\citep{Yang2017b}, and very low metallicity SDSS galaxies \protect\citep{Izotov2019} as different coloured triangles.}
\label{flux_ratios}
\end{figure*}

\section{Results}
The two primary questions that we would like to address with our simulations are: What are the spectral features that differentiate Pop.~III star formation from metal enriched stellar populations and are these signatures detectable? In this Section, we describe the general properties of the simulated galaxy population, discuss the spectral features of these simulated galaxies, and comment on the complexities of identifying a true Pop.~III star or galaxy with observations.

\subsection{High-Redshift Galaxy Properties}
The simulation follows the formation and evolution of the region around a halo with mass $\sim3\times10^{8}$~M$_{\odot}$ by $z=10$. We study the population of haloes inside the high resolution Lagrange region of the main halo. The region contains 356 uncontaminated dark matter haloes with at least 300 particles. In Figure~\ref{hero} we show maps of dark matter, gas, and oxygen, as well as a zoomed in distribution of O{\small I}, O{\small II}, and O{\small III} at $z=10$. At this redshift, the main halo has recently undergone a major merger (a relatively common event at high redshift). There is a significant gas concentration at the centre of the halo and a repeated series of star formation and SN events have enriched a large fraction of the volume with oxygen, some of which has escaped the halo, into the IGM. The structure in the centre of the halo is highly complex and filled with various ionization states of oxygen, dictated by both the local radiation field, SN feedback, and the presence of shocks.

More quantitatively, the first Pop.~III stars formed at $z\sim18$. This can be seen in the top panel of Figure~\ref{sfr_metal_enrich}, where we show the star formation rate as a function of time within the Lagrange region, split between Pop.~III and Pop.~II star formation. Due to the fact that not all Pop.~III stars explode via SN, the first metal enrichment does not occur until $z\sim16.5$. This can be seen as the spike in the O, C, and Fe abundances in the bottom panel of Figure~\ref{sfr_metal_enrich}. The first SN increases the mass-weighted metallicity of the system up to $10^{-4}Z_{\odot}$ in O and Fe and $10^{-5}Z_{\odot}$ in C. The exact enrichment levels for each metal are highly dependent on the mass of the Pop.~III stars that explode. We expect this to vary between haloes at high redshift. 

Once the region becomes metal enriched, Pop.~II star formation can proceed. This is shown as the magenta line in the top panel of Figure~\ref{sfr_metal_enrich}. There is a delay between when the first SN explode ($z\sim16.5$) and when Pop.~II stars form ($z<15$) due to the fact that the gas has to re-settle in the halo that hosted the SN, and for the metals that escaped the halo, it takes time for them to be accreted onto neighbouring systems. A mixed mode of Pop.~III and Pop.~II star formation continues until $z\sim11$. By this time, the mass-weighted metallicity has increased by about a factor of three, at which point star formation is completely dominated by Pop.~II stars. We note that the exact length of the mixed-mode epoch and the redshift at which the Pop.~II-dominant epoch begins are sensitive to both the individual histories of each halo and our definition for the metallicity threshold that separates Pop.~II and Pop.~III stars. If we adopted a higher metallicity threshold, the results here may change. This issue is further discussed below.

Slightly before $z=10$, the halo undergoes a major merger that results in a steep increase in star formation and metal enrichment. By comparing the mass-weighted and volume-weighted gas metallicities in the Lagrange volume, we can analyze the enrichment of the ISM versus the IGM. While the mass-weighted metallicity continuously rises with decreasing redshift, the evolution of volume-weighted metallicity is more complex and sensitive to gas inflowing into the Lagrange region and the ability for stars to eject metals from the haloes. In general, the volume-weighted metallicity remains below the mass-weighted values suggesting that the metals are more concentrated in the haloes rather than in the IGM.

In the absence of heavy elements, H$_2$ is the primary coolant of the ISM. In Figure~\ref{molecule} we show the evolution of total masses of H$_2$, C{\small II} (one of the dominant coolants at higher metallicity), and CO (one of the dominant coolants at high density and high metallicity). At $z\geq16.5$ a baseline level of H$_2$ is formed via reactions with H$^-$. Once the first metals are released into the ISM and IGM, C{\small II} can form as well and simultaneously the main-mode of H$_2$ formation transitions from the primordial channel to formation on dust. Future telescopes such as ngVLA will provide the opportunity to detect C{\small II} in the redshift interval $15\lesssim z\lesssim 20$, which may help probe the properties of the first stars \citep{Carilli2018}. CO is also detectable at high-redshift with the ngVLA, which could also be interesting for placing constraints on the properties of the first stars; however, we find the amount of CO remains limited until sufficient self-shielding can occur in the ISM.

Having described the star formation, metal enrichment, and molecular properties of the gas in the simulation, we continue our analysis by studying the spectral features of the simulated galaxies.

\subsection{The Absence of Metallic Emission Lines}
As Pop.~III stars are by definition metal-free, their star forming clouds are also expected to be of pristine, primordial composition. The observation of a galaxy with strong He~{\small II}~1640\angstrom, H$\alpha$, and H$\beta$ emission with no metal emission lines is potentially a signature of Pop.~III stars \citep{Schaerer2002,Raiter2010,Inoue2011}. From an observational perspective, the absence of metal emission lines will be difficult to prove because the finite integration times and sensitivity limits will only allow for upper limits on metal emission line strengths, rather than robustly proving that metal emission lines are completely absent. The key question is how strong are the metal emission lines expected to be with respect to the Balmer emission, in simulated high-redshift galaxies.

\begin{figure}
\centerline{\includegraphics[scale=1,trim={0 0.0cm 0cm 1cm},clip]{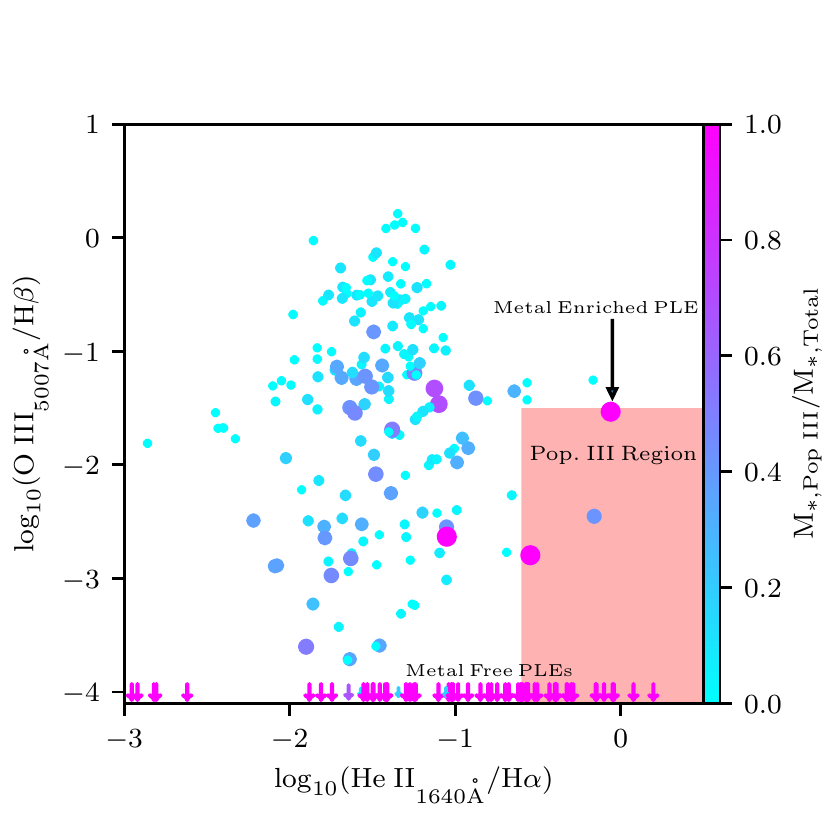}}
\caption{He~{\small II}~1640\angstrom/H$\alpha$ versus [O{\small III}]/H$\beta$ for simulated galaxies in the redshift range $10\leq z\leq16$ that have had star formation in the past 100~Myr and have H$\alpha$ luminosities $>10^{36}\ {\rm erg\ s^{-1}}$. We have coloured the points based on the fraction of the stellar population that is comprised of Pop.~III stars (or their remnants). Downward arrows represent galaxies that have [O{\small III}]/H$\beta<10^{-4}$. The red shaded region in the bottom right represents the parameter space where only Pop.~III-dominated galaxies are found in our simulations. We have labelled the locations of the different types of Pure Line Emitters (PLEs) as discussed in Section~\ref{PLEs}.}
\label{diagnostic_1}
\end{figure}

In Figure~\ref{flux_ratios} we show the flux ratios of [O{\small III}]/H$\beta$ (left), [O{\small II}]/H$\beta$ (centre), and [O{\small III}]/[O{\small II}] (right) for simulated galaxies in the redshift interval $10\leq z\leq16$ coloured by their SFR. Oxygen emission lines are expected to be the brightest metal emission lines at these epochs due to the early enrichment from core-collapse SN \citep[e.g.][]{Maiolino2019}, and the collision strengths of these particular transitions. As the metallicity decreases, [O{\small III}]/H$\beta$ and [O{\small II}]/H$\beta$ also decreases such that at $12+\log_{10}\left(\frac{\rm O}{\rm H}\right)=6$, our simulations predict flux ratios of $\sim10^{-2}$ compared to H$\beta$. Such predictions are not unique to our simulations as {\small CLOUDY} models (see the grey lines and cyan points in Figure~\ref{flux_ratios}) of primordial galaxies predict similar flux ratios \citep{Inoue2011,Nakajima2022}. Considering that our predictions are completely independent of {\small CLOUDY}, it is encouraging that both methods are in reasonable agreement. However, because the simulation samples a diversity of ISM conditions and star formation histories, we do find additional scatter in our relations related to the star formation rate and ISM structure of the galaxy. This is because the star formation rate controls the number of ionizing photons, electron density, and temperature via radiative and SN feedback while the ISM structure dictates how well this feedback couples to the gas. At fixed metallicity, the simulated galaxies exhibit a wide diversity of star formation rates and ISM conditions. Such effects are not captured via simple photoionization models with a fixed uniform gas density and star formation rate.

For comparison, we also show the location of Green Pea galaxies \citep{Yang2017}, Blueberry galaxies \citep{Yang2017b}, and low-metallicity, low-redshift SDSS galaxies \citep{Izotov2019} in Figure~\ref{flux_ratios}. In general, we find good consistency between low-redshift ``analogues'' and our simulated data, providing further confidence in our predictions. The low metallicity galaxies from \cite{Izotov2019} tend to have very comparable [O{\small III}]/H$\beta$, higher [O{\small II}]/H$\beta$, and slightly lower [O{\small III}]/[O{\small II}] compared to the simulated galaxies and {\small CLOUDY} models \citep{Inoue2011,Nakajima2022}. Green Peas and Blueberries tend to be higher metallicity than the simulated galaxies (see also \citealt{Katz2022b}). If we extrapolate the trends from our simulation to higher metallicites, we find no tension between these two galaxy populations and our high-redshift simulations.

\subsection{The Absence of Metal Emission Lines with Strong He~{\small II}~1640\angstrom}
Although we argue that [O{\small III}]/H$\beta$ (or [O{\small II}]/H$\beta$) is, by itself, insufficient to identify a Pop.~III stellar population, if we combine the [O{\small III}]/H$\beta$ ratio with the He~{\small II}~1640\angstrom/H$\alpha$ ratio, we do find a region of parameter space that is only populated by galaxies dominated by Pop.~III stars. The red shaded region in the bottom right of Figure~\ref{diagnostic_1} shows that the parameter space with ${\rm [O{\small III}]/H\beta}<10^{-1.5}$ and ${\rm He~{\small II}~1640\angstrom/H\alpha}>10^{-0.6}$ consists of only systems with stellar populations with a high Pop.~III fraction. Consistent with previous claims in the literature  \citep[e.g.][]{Schaerer2002}, we can confirm that weak metal emission lines and strong He~{\small II}~1640\angstrom~emission is a signature of Pop.~III stars in our simulations.

Within this Pop.~III parameter space, we find three interesting features. There are numerous systems that have [O{\small III}]/H$\beta=0$ (represented as downward arrows in Figure~\ref{diagnostic_1}). These are genuine Pop.~III stars forming in pristine, zero-metallicity gas clouds. These types of systems are what is typically expected of Pop.~III stars. 

\begin{figure}
\centerline{\includegraphics[scale=1,trim={0 0.0cm 0cm 0cm},clip]{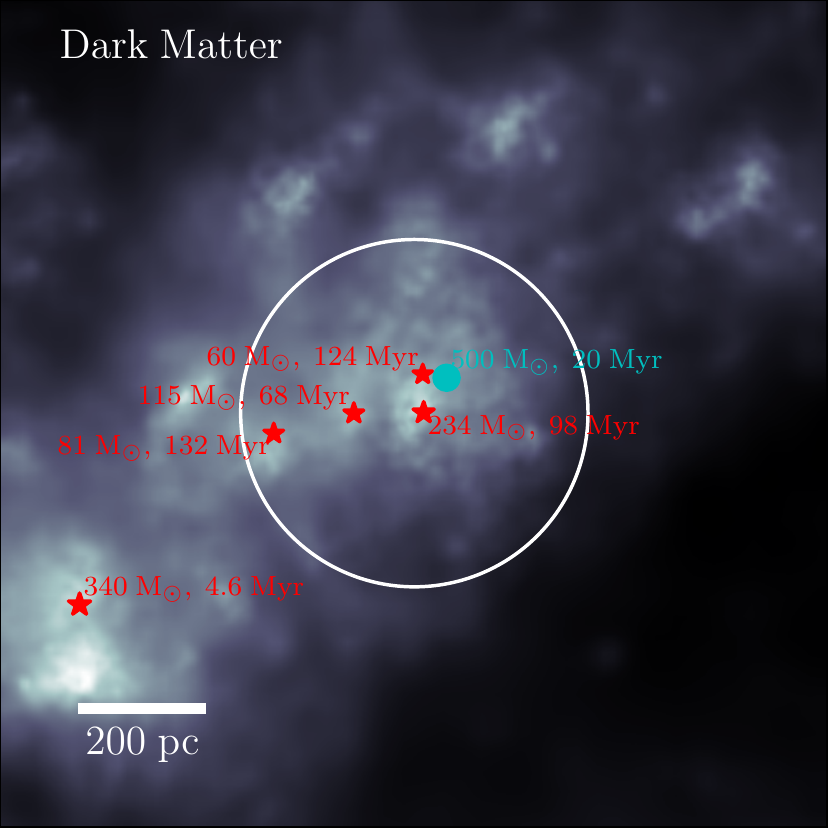}}
\caption{Dark matter map of a galaxy with mixed Pop.~III and Pop~II star formation at $z=12.6$, The red stars represent Pop.~III stars and the cyan circle shows the location of a Pop.~II star cluster. The ages and initial masses of each star particle are annotated on the map. The circle represents the virial radius of the halo (588~pc).}
\label{mixed_map}
\end{figure}

Next we have a galaxy that appears as the blue point in the shaded region that hosts a stellar population that is a mix of Pop.~II and Pop.~III stars. Such mixes are common in our simulation, although not all have high He~{\small II}~1640\angstrom~emission. A dark matter map of this galaxy with the locations of the Pop.~II and Pop.~III star particles is shown in Figure~\ref{mixed_map}. These mixed systems add to the complexity of detecting a fully Pop.~III dominated galaxy. Simulations with larger volumes may find similarly complex systems of this nature that potentially pollute the region of Figure~\ref{diagnostic_1} that we ascribe as being Pop.~III. 

Going into more detail on this specific galaxy, within 100~pc of the centre, we find two Pop.~III star particles of ages 98~Myr and 124~Myr, the former of which exploded as a Pair-Instability SN (PISN) and the latter collapsed directly into a black hole. The third star particle is a 500~M$_{\odot}$ cluster of Pop.~II stars with an age of 20~Myr that is driving most of the line emission. This system is surrounded (within 2~kpc) by three more Pop.~III stars, one of which is 340~M$_{\odot}$, with an age of 4.6~Myr, that is likely helping to drive some of the He{\small II} emission externally. 

It is also important to consider that the origin of strong He{\small II} emission is unknown at low redshifts and it is possibly driven by X-ray binaries, Wolf-Rayet stars, or other sources \citep[e.g.][]{Erb2010,Schaerer2019}. If our chosen BPASS SED neglects these sources, we are potentially underpredicting the He{\small II} luminosities of our Pop.~II galaxies. This would further pollute this diagnostic, making it more difficult to identify genuine Pop.~III stars. 

\begin{figure}
\centerline{\includegraphics[scale=1,trim={0 0.0cm 0cm 0cm},clip]{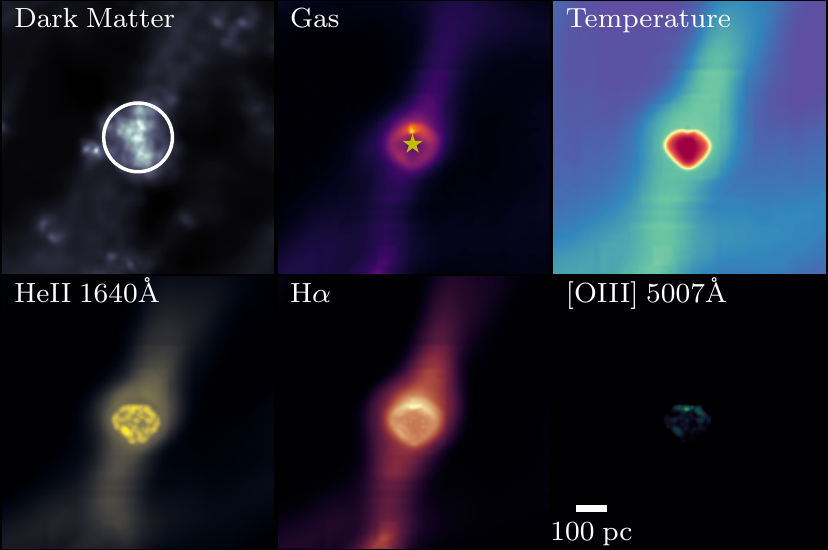}}
\caption{Maps of the dark matter, gas, temperature, He{\small II}~1640\angstrom, H$\alpha$, and [O{\small III}]~5007\angstrom~are shown for a Pop~III SN remnant that had an initial mass of 180.59~M$_{\odot}$. The circle in the top left panel represents the virial radius of the halo (292~pc) and the yellow star in the top centre panel shows the location of the star particle when the SN exploded.}
\label{snremnant}
\end{figure}

Finally, we find galaxies that are fully dominated by Pop.~III stars but exhibit metal emission because we are observing a cooling SN remnant. Hence the presence of metal emission lines does not necessarily rule out the system from being Pop.~III, even though this cooling phase is expected to be short-lived.

More specifically, there is a galaxy that is completely dominated by Pop.~III stars but has [O{\small III}]/H$\beta=10^{-1.5}$. We show maps of the dark matter, gas, and temperature, and the surface brightness maps of He{\small II}~1640\angstrom, H$\alpha$, and [O{\small III}]~5007\angstrom~for this galaxy in Figure~\ref{snremnant}. This system hosted a single 180~M$_{\odot}$ Pop.~III star that recently exploded as a PISN. The PISN enriched the surrounding medium with metals so the emission that is seen is a combination of the recombination emission due to photoionization from the star and the cooling radiation from the SN remnant. The interesting characteristic of this galaxy is that there is no stellar continuum. By definition (excluding the contribution from nebular continuum), the equivalent widths of all of the emission lines from this system are infinity. We discuss such systems further in Section~\ref{PLEs}. If another Pop.~III star formed in this system before the first evolved off the main-sequence, we might expect an increase in luminosity for all emission lines, including those from metals. 

\begin{figure}
\centerline{\includegraphics[scale=1,trim={0 0.0cm 0cm 0cm},clip]{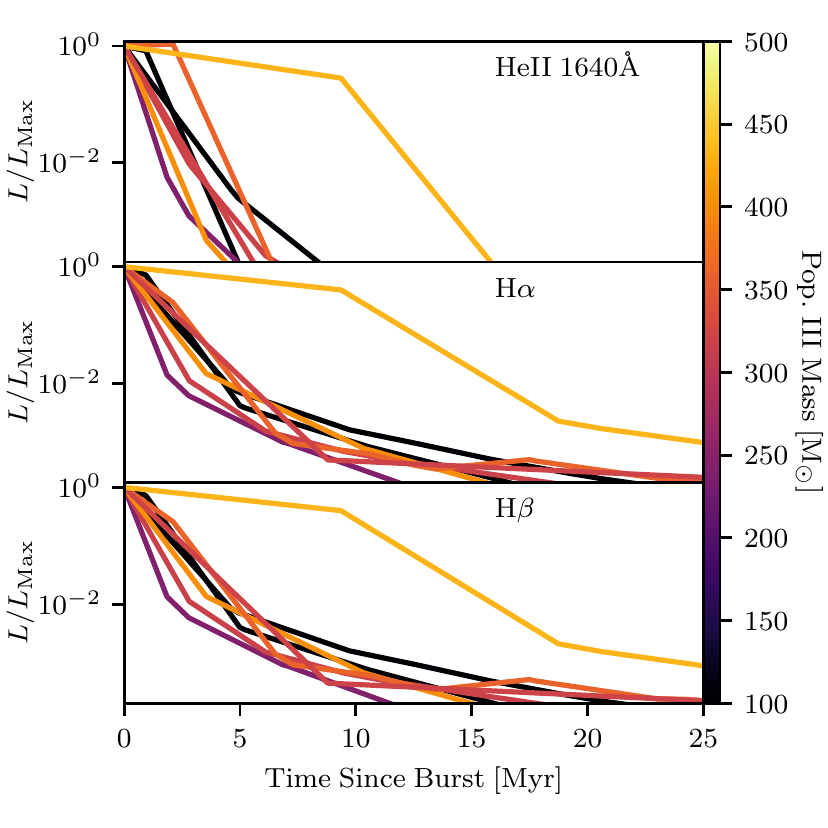}}
\centerline{\includegraphics[scale=1,trim={0 0.0cm 0cm 0cm},clip]{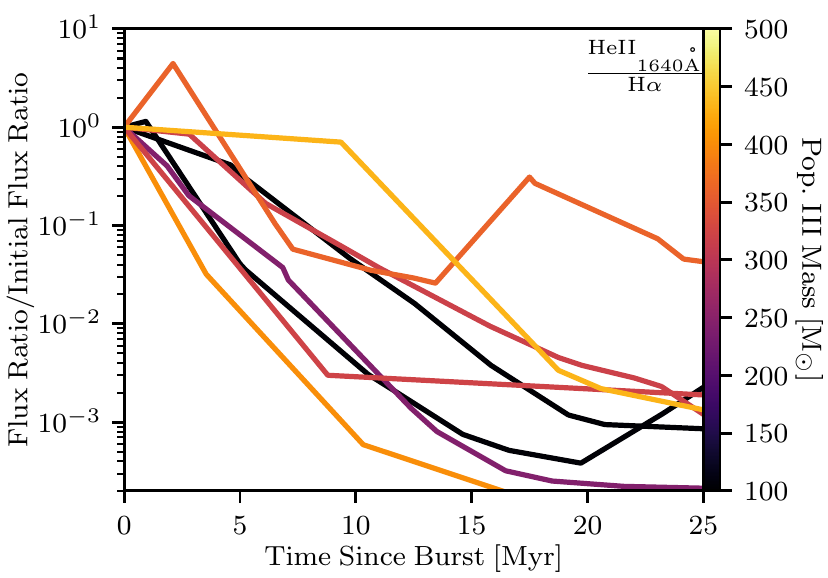}}
\caption{(Top) Time evolution of He{\small II}~1640\angstrom, H$\alpha$, and H$\beta$ emission normalized to the maximum luminosity of each emission line after the burst as a function of time since the formation of a Pop.~III star for eight halos in the simulation. The colour of the line indicates the mass of the Pop.~III star that formed during the initial burst. (Bottom) Flux ratio of He{\small II}~1640\angstrom~to H$\alpha$ normalized by the ratio at the time of the burst. In most cases, the emission line strengths fade within 5~Myr, with He{\small II}~1640\angstrom~fading faster than the Balmer emission. }
\label{lifetime}
\end{figure}

\subsection{Time Windows for Detection}
If Pop.~III stars were predominantly massive, their main-sequence lifetimes would be very short. Hence the emission line signatures are also expected to be short lived \citep[e.g.][]{Schaerer2002,Schaerer2003}. Therefore, Pop.~III lifetime is an important parameter when considering their detectability \citep{Zackrisson2012,Rydberg2013}.

In the top panels of Figure~\ref{lifetime}, we show the time evolution of the He{\small II}~1640\angstrom, H$\alpha$, and H$\beta$ emission for eight haloes in the simulation as a function of time after the formation of a Pop.~III star. The lines are coloured based on the initial mass\footnote{In the case where multiple Pop.~III stars form, the line is coloured based on the mass of the first star to form.} of the Pop.~III star that formed in the halo. The luminosities have been normalized to the maximum luminosity of the line recorded after the star formation event\footnote{Note that because the simulation outputs snapshots at fixed times, there is often some delay after the formation of each star before we can first measure the emission line luminosities. The time cadence of our snapshots is often $<5$~Myr so we expect our curves to be reasonably time-resolved.}. In most cases, the He{\small II}~1640\angstrom~fades by nearly two orders of magnitude within 5~Myr and the Balmer emission shows similar properties. This result is unsurprising given that our Pop.~III stars are in general massive and we have adopted the SED of \cite{Schaerer2002}. The unique aspect of this work is that we can see how this signal fades for a realistic gas distribution. In the bottom panel of Figure~\ref{lifetime}, we show the ratio of He{\small II}~1640\angstrom/H$\alpha$ normalized to the value of this ratio after the star formation event. In general, the He{\small II}~1640\angstrom~emission decreases faster than H$\alpha$. This explains why the Pop.~III galaxies in Figure~\ref{diagnostic_1} that have no metal emission lines also show weak He{\small II}~1640\angstrom/H$\alpha$.

There is a single halo in our simulation where the He{\small II}~1640\angstrom~emission remains strong for $\sim10$~Myr. Looking at this system in more detail, we find that a second Pop.~III star formed in the galaxy very soon after the first. The presence of the second Pop.~III star maintains the ionizing luminosity needed to continually excite the He{\small II}~1640\angstrom~line. This phenomenon is considerably rarer in our simulation than isolated Pop.~III star formation, but we note that these predictions are highly IMF/model dependent. Nevertheless, for a top-heavy Pop.~III IMF, the number densities of galaxies with He{\small II}~1640\angstrom~emission may be lower than that for a more bottom-heavy IMF.

It is important to consider that the formation of a single massive Pop.~III star can enrich a galaxy up to the critical metallicity needed for Pop.~II star formation in our model. As the galaxies become more massive, it can maintain a higher star formation rate and sustain bright emission line luminosities for a much longer period of time. Furthermore, our chosen Pop.~II SED ({\small BPASS}) allows for ionizing photons for $>10$~Myr after the formation of a star particle, which represents a much longer window for detection of a single burst compared to a Pop.~III stellar population. If all haloes that produce Pop.~III stars eventually go on to form Pop.~II stellar populations, the ratios of time window for detection means that the vast majority of detected systems will be Pop.~II.

\subsection{Pure Line Emitters as a Signature of Pop.~III Stars}
\label{PLEs}
Due to the nature of our chosen Pop.~III IMF, the stars can either end their lives as normal type-II SN, PISN, or by directly collapsing into black holes. In all three cases the gas in the host halo will emit for some period of time after the Pop.~III star has evolved off the main sequence. In this case, the galaxy may appear to have strong emission lines (and possibly a nebular continuum), but no stellar continuum due to the fact that the star formed in isolation. We refer to these galaxies as pure line emitters (PLEs), although similar signatures have been found for AGN \citep[e.g.][]{Lintott2009}.

We find two types of PLEs in the simulation: metal-enriched PLEs, which result from a single Pop.~III SN, and metal-free PLEs that are generated by the formation of a Pop.~III star that subsequently collapses directly into a black hole. Both are labelled in Figure~\ref{diagnostic_1}. The two types of PLE have different emission line signatures (i.e. the presence of metals or not), but both seem to have emission lines that fade very fast. The gas density at which the SN explodes as well as the gas distribution \citep{Blondin1998,Thornton1998} may impact the rate at which the SN remnant cools; thus the single PISN-driven PLE in our simulation may not be fully representative of the expected parameter space. Nevertheless, detecting these objects may be difficult due to their short-lived nature, but if equivalent widths can be constrained, they may provide an alternative metric for identifying Pop.~III systems.

It is important to consider that there may be other types of objects that can be confused with PLEs. For example, a black hole that turns on and off in a pristine or even metal enriched galaxy could potentially show similar features to a PLE. \cite{Nakajima2022} showed that the He{\small II}~1640\angstrom/Ly$\alpha$ ratio (and therefore also He{\small II}~1640\angstrom/H$\alpha$) is similar between Pop.~III stars and direct collapse black holes, peaking at a few percent. Only the equivalent width of He{\small II}~1640\angstrom~could differentiate the two types of sources (when only these lines are observed). If the source turns off, the equivalent width cannot be measured. If star formation is very heavily dust obscured than the continuum may not be detectable, but if there is a gas cloud near the star-forming region that is not obscured, this could mimic a PLE. However, we would not expect hard spectral features unless the system contained, for example, X-ray binaries, Wolf-Rayet stars, a black hole, etc. We would not expect this setup to be confused with our metal-free PLEs unless the nearby gas cloud is also pristine or of very low metallicity. 

\section{Caveats}
The primary systematic uncertainties with our work are the choices of Pop.~III IMF, main-sequence lifetimes, SED, post main-sequence behaviour, metal yields, and the transition metallicity between Pop.~III and Pop.~II star formation. Ideally we could run more simulations to vary these parameters; however, the computational expense of these simulations poses a severe limitation. Nevertheless, one can speculate how changes to these assumptions will impact our results.

For example, moving to a less top heavy IMF may allow the spectral signatures of Pop.~III stars to last much longer due to the extended main-sequence lifetimes. Making the SEDs less or more hard will impact the strength of the He{\small II}~1640\angstrom~line with respect to the Balmer lines by making the ratios weaker or stronger, respectively. Depending on whether the Pop.~III stars collapse directly to black holes or explode as SN, this could change the timescale for enrichment. The abundance patterns of the enriched gas will change depending on the ratio of PISN to normal core-collapse SN. Given that our models are in good agreement with results from previous {\small CLOUDY} models \citep{Inoue2011,Nakajima2022}, barring the additional scatter due to SFR, galaxy structure, etc., varying these parameters in photoionization models will provide a good estimate for the expected signatures of Pop.~III stars. 

One of the key uncertainties with our modelling is whether Pop.~III stars form in isolation or in large groups. Due to resolution, most Pop.~III stars in our work form in isolation; however, higher resolution simulations that better resolve the star formation process show that Pop.~III stars can form in binaries or small groups \citep[e.g.][]{Stacy2016}. This could help increase the time period over which He{\small II} is bright.

Due to computational expense, we have only run one simulation of the Lagrange region around a single halo. As described earlier, the onset of Pop.~III star formation and length of the Pop.~III-Pop.~II transition will be halo and environment dependent. Nevertheless, we expect qualitatively similar behaviour in other environments. The galaxies in our simulation are intrinsically faint, but we expect the emission line ratios to hold for any more massive galaxy with similar ISM properties that may be more luminous. More simulations will be required to sample the true diversity of high-redshift galaxies.

\section{Conclusions}
We have run a high-resolution cosmological radiation hydrodynamics zoom-in simulation of the region around a dwarf galaxy at $z\geq10$ with sub-pc resolution to predict the emission line signatures of the Pop.~III/Pop.~II transition. The simulations are uniquely run with the {\small RAMSES-RTZ} code \citep{Katz2022}, which allows us to predict emission line signatures from both primordial species and metals in a fully non-equilibrium manner.  

We predict that early metal enrichment from Pop.~III stars can quickly increase the mass-weighted metallicity of the gas up to $10^{-4}Z_{\odot}$ as measured by O and Fe abundances, which is similar to the results of other simulations \citep[e.g.][]{Wise2012c}. This metallicity floor is enough to initiate the formation of the first generations of Pop.~II stars at $z\sim16.5$. A mixed-mode of Pop.~III and Pop.~II star formation continues until $z\sim11$, when the Lagrange region is sufficiently metal enriched enough so that Pop.~II star formation dominates the volume.

We calculate emission line luminosities of [O{\small III}]~5007\angstrom, [O{\small II}]~3727\angstrom, H$\alpha$, H$\beta$, and He{\small II}~1640\angstrom~and show that even among metal enriched galaxies (i.e. those with $Z\gtrsim0.01Z_{\odot}$), the  [O{\small III}]~5007\angstrom/H$\beta$ ratio can remain below $10^{-2}$, consistent with {\small CLOUDY} calculations \citep{Inoue2011,Nakajima2022}, indicating that it will be difficult to identify a truly metal free galaxy with JWST due to sensitivity limits. Combining this ratio with He{\small II}~1640\angstrom/H$\alpha$, we find that there is parameter space with ${\rm [O{\small III}]/H\beta}<10^{-1.5}$ and ${\rm He~{\small II}~1640\angstrom/H\alpha}>10^{-0.6}$ that is populated only by galaxies dominated by Pop.~III star formation. However, this parameter space is sparsely populated because the He{\small II}~1640\angstrom~emission fades very quickly (i.e. within a $\sim5$~Myr) after the onset of star formation. Hard He{\small II}~1640\angstrom/H$\alpha$ ratios can only be maintained for longer periods of time if multiple Pop.~III stars form in sequence or with different masses in the same system, which is rare in our model. Finally, we show that galaxies with emission lines but no detectable continua (i.e. PLEs) could offer a signature of a Pop.~III stars that either subsequently collapsed directly into a black hole (if there are no metal emission lines) or Pop.~III stars that exploded via SN (if there are metal emission lines). These ``infinite'' equivalent width galaxies are also expected to be short lived, but could represent a further diagnostic of Pop.~III star formation if detected.

In summary, JWST will provide an exciting probe of star formation and possibly even Pop.~III star formation in the early Universe. However, our work demonstrates that there are significant challenges in robustly identifying a Pop.~III stellar population that should be taken into consideration when analysing upcoming JWST data.

\section*{Acknowledgements}
HK thanks Aayush Saxena, Martin Rey, Alex Cameron, Eric Andersson, Oscar Agertz, and Girish Kulkarni. This project has received funding from the European Research Council (ERC) under the European Union’s Horizon 2020 research and innovation programme (grant agreement No 693024). TK was supported by the National Research Foundation of Korea (NRF) grant funded by the Korea government (No. 2020R1C1C1007079 and No. 2022R1A6A1A03053472). RSE acknowledges funding from the European Research Council under the European Union Horizon 2020 research and innovation programme (grant agreement No. 669253). Some of this work used the DiRAC@Durham facility managed by the Institute for Computational Cosmology on behalf of the STFC DiRAC HPC Facility (www.dirac.ac.uk). The equipment was funded by BEIS capital funding via STFC capital grants ST/P002293/1, ST/R002371/1 and ST/S002502/1, Durham University and STFC operations grant ST/R000832/1. Some of this work was performed using the DiRAC Data Intensive service at Leicester, operated by the University of Leicester IT Services, which forms part of the STFC DiRAC HPC Facility (www.dirac.ac.uk). The equipment was funded by BEIS capital funding via STFC capital grants ST/K000373/1 and ST/R002363/1 and STFC DiRAC Operations grant ST/R001014/1. DiRAC is part of the National e-Infrastructure.

\section*{Data Availability}
The data underlying this article will be shared on reasonable request to the corresponding author.

\bibliographystyle{mnras}
\bibliography{example}




\bsp	
\label{lastpage}
\end{document}